# Non-linear second-order topological insulators

Farzad Zangeneh-Nejad and Romain Fleury*

*Laboratory of Wave Engineering, School of Electrical Engineering, EPFL, Station 11, 1015 Lausanne, Switzerland*

*To whom correspondence should be addressed. Email: romain.fleury@epfl.ch

**We demonstrate, both theoretically and experimentally, the concept of non-linear second-order topological insulators, a class of bulk insulators with quantized Wannier centers and a bulk polarization directly controlled by the level of non-linearity. We show that one-dimensional edge states and zero-dimensional corner states can be induced in a trivial crystal insulator made of evanescently coupled resonators with linear and nonlinear coupling coefficients, simply by tuning the excitation intensity. This allows global external control over topological phase transitions and switching to a phase with non-zero bulk polarization, without requiring any structural or geometrical changes. We further show how these non-linear effects enable dynamic tuning of the spectral properties and localization of the topological edge and corner states. Such self-induced second-order topological insulators, which can be found and implemented in a wide variety of physical platforms ranging from electronics to microwaves, acoustics, and optics, hold exciting promises for reconfigurable topological energy confinement, power harvesting, data storage, and spatial management of high-intensity fields.**

Topological insulators (TIs) are recently discovered phases of matter that, despite their d-dimensional insulating bulk behavior, host $d-1$ dimensional gapless conductive states at their boundaries. The presence of these edge modes is guaranteed and protected by the topological property of the surrounding bulk insulators, which is characterized by a quantized number known as topological invariant [1-6]. The protected nature of the topological edge modes has offered the unique possibility of robust 1D transport of information and power in various fields of interest including quantum mechanics [7,8], optics [9-29], acoustics [30-42], microwaves [43-46], and mechanics [47,48].

More recently, the conventional classification of topological phases has been generalized by introducing a novel form of symmetry-protected topological systems hosting resonant states not only on their edges, but also on the edges of their edges [49]. In particular, in the 2D case, these so-called second-order topological insulators (SOTIs) [50-54], characterized by quantized Wannier centers [49], support zero dimensional (0D) corner states that are bounded to the extremities of the insulator edges. The strong confinement of these 0D topological states represents a disruptive advance in our ability to build reliable energy storage and harvest points, which is relevant in various application areas ranging from silicon nanophotonics to mechanical energy management.

So far, demonstrations of SOTIs in electronic [55], microwave [56], and acoustic [57,58] systems have been restricted to linear structures, where non-linearities are absent. Recent works, however, have suggested non-linear phenomena as an important tool for achieving ordinary TIs. In particular, it has been shown [59,60] that a non-linear Su-Schrieffer–Heeger (SSH) chain can support distinct topological phases that are controlled by the intensity level, with its bulk modes turning into topological edge modes after passing a certain intensity threshold. In comparison with their linear counterparts, such kinds of topological edge states have two important advantageous properties: (i) They do not imply any stringent structural design as they can be externally induced by enhancing the excitation intensity, or equivalently, the level of non-linearity; (ii) Their spectral properties and localization can be dynamically controlled by tuning the excitation power. Despite these advances, the extension of first-order nonlinear TIs to their second-order counterparts has remained largely uncharted. In particular, nonlinear effects might be leveraged as an inherent degree of freedom to quantize the Wannier centers of a trivial insulator and achieve self-induced higher order topological phases supporting 0D corner states.

Here, we demonstrate the concept of nonlinear SOTIs, in which Wannier centers become quantized by enhancing the excitation power. More specifically, we show how a topologically trivial tight-binding lattice with linear intra-cell and nonlinear extra-cell coupling coefficients can be turned into a second-order topological insulator, supporting self-induced zero-dimensional corner modes. We further prove that the characteristics of the corresponding topological corner states, such as their confinement and spectrum, can be dynamically tuned by adjusting the excitation intensity. We finally report an experimental validation of these findings in a non-linear circuit lattice. Considering the generality of our approach, based on the standard tight-binding formalism, these self-induced and reconfigurable SOTIs can be found in other applicative fields, including photonics, phononics, and microwaves, opening a wealth of new opportunities for topological energy confinement and data storage.

We first discuss the core idea of our proposal, whose concept is sketched in Fig. 1. Consider a $C_6$ symmetric piece of crystal, made of a contracted hexagonal lattice of evanescently-coupled resonators (panel a) with linear intra-cell (dashed blue lines) and nonlinear extra-cell (dashed red lines) coupling coefficients. Here, the term "contracted" refers to the fact that the unit cell is constructed starting from a regular honeycomb lattice and contracting the size of its hexagons (in [61] we consider another example with a different lattice). For the purpose of conceptual demonstration, we can for instance think of an optical system that is probed, for instance, from the far field (as we discuss below, other physical platforms and form of excitation are also possible, as in our experiment). Suppose first that the crystal is probed with a low intensity pump so that the non-linear effects are negligible. Such a linear contracted honeycomb lattice supports a complete band gap [62,63], which is topologically trivial and corresponds to a vanishing bulk polarization. Consequently, no states are supported at the edges or corners of the crystal.

To induce a topological phase transition in such a system, a possibility is to increase the strength of the extra-hexagon coupling until it becomes larger than the intra-cell one, closing the trivial band gap and re-opening it as topological with a non-vanishing bulk polarization, characterized by quantized Wannier centers (see [61]). While this SOTI could be realized by expanding the size of the hexagon within the unit cell of the crystal, an elegant alternative is to leverage the inherent nonlinearities of the system to trigger a topological transition without changing the lattice geometry. As schematically shown in Fig. 1b, we propose to achieve this just by enhancing the excitation intensity.

More formally, we express the dynamics of the lattice under study with a non-linear Schrodinger equation [61] as

$$\widehat{H}_{|\mathbf{m}\rangle}|\mathbf{m}\rangle = (\omega - \omega_0 \mathbf{I})|\mathbf{m}\rangle \qquad (1)$$

where $|\mathbf{m}\rangle$ is the eigenvector, $\omega_0$ is the resonance frequency of individual resonators, $\mathbf{I}$ is the identity matrix, $\omega$ is the eigenvalue, and $\widehat{H}_{|\mathbf{m}\rangle}$ is the tight-binding Hamiltonian, which can, in the framework of coupled mode theory, be written as

$$\widehat{H}_{|\mathbf{m}\rangle} = \sum_m \widehat{K}|\mathbf{m}\rangle\langle\mathbf{m}| + \sum_{m,n}(\hat{J}_{m,n}|\mathbf{m}\rangle\langle\mathbf{n}| + \hat{J}^*_{m,n}|\mathbf{n}\rangle\langle\mathbf{m}|) \qquad (2)$$

In (2), $\widehat{K}$ and $\hat{J}_{m,n}$ are matrices describing the intra-cell and extra-cell coupling (blue and red links in Fig. 1, respectively). Note that $\hat{J}_{m,n}$ is assumed to have dependency on the norm of the eigenvectors $|\mathbf{m}\rangle$ and $|\mathbf{n}\rangle$, thus the eigenvalue equation (1) should be solved iteratively (see [61]). Here, we assume a Kerr-like nonlinearity for the extra-cell dynamics, with $\hat{J}_{m,n} = \hat{J} + \hat{\alpha}_{m,n} I$, where $\hat{\alpha}_{m,n}$ and $I$ are the nonlinear Kerr matrix and the total on-site intensity, respectively. Assuming that the nonlinearity is weak enough so that the modal solution can be approximated using a Bloch-like wave function, one can derive the kernel of Hamiltonian in momentum space as [61]

$$\widehat{H} = \lambda(\hat{\sigma}_s^{1,6} + \sum_{i=1}^{5} \hat{\sigma}_s^{i,i+1}) + \sum_{i=1}^{3} A_{i,i+3} \hat{\sigma}_s^{i,i+3} + \sum_{i=1}^{3} B_{i,i+3}\, \hat{\sigma}_a^{i,i+3}) \qquad (3)$$

where $\hat{\sigma}_s^{i,j}$ and $\hat{\sigma}_a^{i,j}$ are the symmetric and asymmetric Gell-Mann matrices, respectively, $\lambda$ is the intracell coupling coefficient and $A_{i,i+3}$ and $B_{i,i+3}$ are of the form

$$A_{i,i+3} = \big(\Lambda + \alpha I(|m_i|^2 + |m_{i+3}|^2)\cos(\boldsymbol{k_B} \cdot \boldsymbol{r}_{i,i+3})\big)$$

$$B_{i,i+3} = \big(\Lambda + \alpha I(|m_i|^2 + |m_{i+3}|^2)\sin(\boldsymbol{k_B} \cdot \boldsymbol{r}_{i,i+3})\big) \qquad (4)$$

in which $\Lambda$ is the intercell coupling coefficient, $\alpha$ is the Kerr non-linear coefficient, $\boldsymbol{k_B} = k_x \hat{\boldsymbol{x}} + k_y \hat{\boldsymbol{y}}$ is the Bloch wave number, $\boldsymbol{r}_{14} = a/2\hat{\boldsymbol{x}} - a\sqrt{3}/2\,\hat{\boldsymbol{y}}$, $\boldsymbol{r}_{25} = a\hat{\boldsymbol{x}}$, and $\boldsymbol{r}_{36} = a/2\,\hat{\boldsymbol{x}} + a\sqrt{3}/2\hat{\boldsymbol{y}}$, with $a$ being the lattice constant. We arbitrarily choose the values of intracell, inter cell couplings, and Kerr coefficient, to be $\lambda = 2.5 \times 10^{-3}$, $\Lambda = 2.3 \times 10^{-3}$ and $\alpha = 5 \times 10^{-5}$. We first focus on the weak intensity case where the non-linear effects are small. Fig. 2a reports the calculated band structure of the lattice crystal (along $\Gamma K$ direction) in case of infinitely small non-linearities. As seen in the figure, there exists a band gap, marked with the purple color. As we will see in the following, this insulating phase, which is the same as that of a linear contracted hexagonal crystal [57], is topologically trivial, and has a vanishing bulk polarization.

We now increase the intensity. The frequency band gap then starts shrinking. At some threshold intensity level $I_{th}$, the band gap completely closes (Fig. 2b), creating a fourfold degenerate point at $\Gamma$. Such band structure indeed resembles that of an undeformed hexagonal lattice with two folded Dirac cones at the center of the Brillouin zone. When the intensity is increased further (Fig. 2c), the Dirac degeneracies are lifted, reopening a band gap, with a potentially different topology.

To evaluate whether the regimes before and after the threshold intensity correspond to distinct topological phases, we calculate the bulk polarization of the crystal, defined by $P = \frac{1}{N}\sum_{j=1}^{N} \vartheta^j$, where $\vartheta^j$ are the Wannier centers, and $N$ is their total number. The Wannier centers $\vartheta^j$ are the eigenvalues of the Wilson loop problem [61]. Shown in Fig. 2d is the evolution of the polarization $P$ versus excitation intensity. As observed, the bulk polarization jumps from zero to $1/2$ when crossing the threshold intensity $I_{th}$, clearly revealing that the non-linearities drive a topological phase transition at threshold. We note that the obtained quantized values of bulk polarization (i.e. $P = 0, 1/2$) are consistent with what is expected from symmetry arguments (see [61]).

We now focus our attention on the associated zero-dimensional corner states. To this end, we consider a finite $C_6$ symmetric sample of the previous non-linear honeycomb lattice, composed of 271 unit cells arranged in a hexagonal shape. Fig. 3a represents the corresponding eigenvalue distribution as the intensity level increases. As anticipated, at low excitation levels (say for example at $I_1$), there exists a set of bulk Bloch modes (in grey) separated by a band gap. However, when the excitation power is increased ($I_2$), the band gap closes as a direct consequence of the topological phase transition described above, creating topological states that, according to their profiles, are classified into two groups: (i) Ordinary helical edge modes (in blue) and (ii) Second-order zero-dimensional corner modes (in red). These observations broadly demonstrate how non-linearities can be leveraged to realize a SOTI without any lattice deformation. Note that the topological corner states are protected by the $C_6$ symmetry of the lattice and remain intact as long as this symmetry is preserved and the bandgap is open [61].

Not only do the non-linearities allow one to tailor the topological phase of the system, but they also establish a unique platform for tuning the spectral properties of the edge and corner modes. To assess this remarkable possibility, we calculate the spectrum of the density of states at three different excitation levels, namely $I_1$, $I_2$, and $I_3$, and plot the results in Figs. 3b to d, respectively. As observed, when the excitation level is small (panel b), neither a corner nor an edge mode appears in the resolved spectrum, since the system is topologically trivial. At the excitation levels $I_2$, and $I_3$, on the contrary, the system becomes topological and hosts edge and corner modes, which are colored according to our previous convention (panel c and d). The spectrum of the crystal is however not the same for the latter two intensities despite the fact that they both correspond to the same nontrivial topological phase. More specifically, the frequency band gap is larger for the higher excitation intensity ($I_3$), allowing a more localized realization of the edge and corner states.

We next report in Fig. 4 the spatially resolved energy distribution of a bulk (column a), edge (column b) and corner (column c) mode at the three specified intensity levels, i.e. $I_1$, $I_2$, and $I_3$ (tracking specific bands). For the intensity level $I_1$, corresponding to a trivial topology, the structure only supports ordinary Bloch states living in the bulk of the crystal. For the intensity level $I_2$, however, the crystal becomes topological and supports both helical edge modes and corner states. When the intensity is further enhanced to $I_3$, the topological states become more confined. This higher confinement is consistent with the larger spectral band gap that the crystal exhibits for higher excitation intensities. This observation suggests the interesting possibility of dynamically tuning the localization of the corner modes (and also the edge modes), simply by changing the excitation power.

Based on our theoretical findings, we designed and built a prototype and experimentally demonstrated the emergence of nonlinearity-induced 0D corner modes. We mapped our tight binding model into the electronic circuit depicted in Fig. 5a, in which the atoms of the tight binding lattice are replaced with *LC* tanks (see [61] for details). The *LC* resonators inside each hexagon are coupled to each other via linear capacitors (red), whereas the nearest neighboring cells are connected to each other through back-to-back varactor diodes [60,64,65] (blue), effectively creating an intensity-dependent capacitive coupling between the adjacent cells. We note that the nonlinearity that the varactor diodes provide is different than a Kerr-type nonlinearity, yet without changing the physics. At low excitation power $P_{in}$, the circuit is designed to be topologically trivial with its intra-cell couplings being larger than the extra-cell ones. The presence of nonlinear varactor diodes, however, gives rise to an enhanced extra-cell coupling at higher intensities, allowing us to reach the SOTI phase. To demonstrate this, we numerically calculate the spectrum of the input admittance of the circuit when it is excited from its corner at $P_{in} = 0\ dBm$, corresponding to a trivial topology (Fig. 5b (top)). The spectrum represents multiple resonant peaks, corresponding to bulk modes. Conversely, upon increasing $P_{in}$ to 25 *dBm* (Fig. 5b, bottom), a clearly dominant resonance appears in the spectrum. This resonance corresponds to the topological corner state, whose intensity profile is numerically extracted and illustrated in Fig. 5c. A picture of our prototype is shown in Fig. 5d. The measured input admittance spectrum and intensity profile of the corner mode are shown in Figs. 5e,f, respectively. Our numerical predictions and experimental measurements are in full agreement.

In summary, this work proposed the concept of non-linear second-order topological insulators, which are bulk insulators supporting intensity-dependent quantized Wannier centers. We showed how a trivial tight-binding lattice with linear intra-cell and nonlinear extra-cell coupling coefficients can be turned into a topological one, hosting both ordinary helical edge modes and corner modes, which we experimentally observed in a circuit prototype. One can also envision similar implementations of such topological phases of matter, for instance, in optics using an array of nanoparticles or high contrast dielectrics [66-68], connected to each other via linear and nonlinear dielectrics. We expect this work to trigger significant developments in the study of non-linear topological phases in systems not only with cubic non-linearities, but also with non-linearities of different origin. Finally, while we restricted ourselves to the small signal regime, the study of second order topological insulators in the high-oscillation regime and the analysis of their stability seem like a promising research direction [69,70].

**Figures**

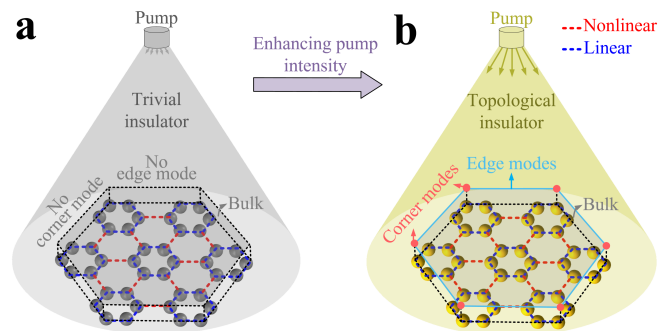

**Fig. 1: Concept of non-linear second-order topological insulators (SOTIs).** A tight-binding lattice made from dielectric resonators, with linear (blue) and nonlinear (red) nearest-neighbor couplings. **a,** When the excitation intensity is low, the structure behaves as a linear crystal, which is a trivial topological insulator: it supports bulk modes, but no edge or corner states. **b,** By increasing the intensity of the pump until it goes above a certain threshold, a topological phase transition occurs and the initial trivial insulator is turned into a SOTI, hosting both edge and corner states.

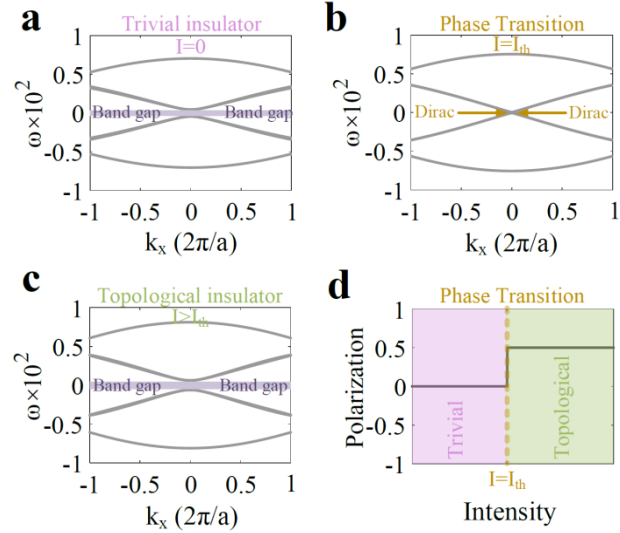

**Fig. 2: Band structure of the non-linear insulator for different intensities. a,** Band structure of the non-linear crystal when the excitation intensity is small. Similar to the linear case, the band structure possesses a band gap, which is topologically trivial. **b,** Band structure at the threshold intensity $I_{th}$, at which the band gap closes, creating fourfold degenerated modes at the center of the Brillouin zone. **c,** Increasing the intensity level further, the band gap re-opens with a non-trivial topology. **d,** Evolution of the bulk polarization of the lattice crystal versus intensity. The bulk polarization jumps from 0 to 0.5 when crossing the threshold intensity level, revealing the topological phase transition occurring at this intensity.

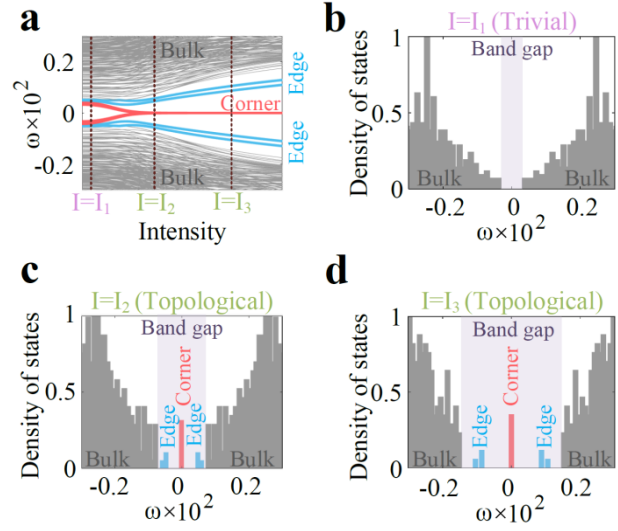

**Fig. 3: Demonstration of self-induced edge states in a non-linear SOTI. a,** Evolution of the eigenfrequency spectrum of a finite hexagonal piece of shrunken hexagonal lattice, composed of 271 unit cells, as the intensity level is increased. At low intensity levels ($I_1$), the crystal supports a set of bulk Bloch modes (in grey) separated by a band gap. When the excitation is enhanced, the band gap closes through a topological phase transition, creating ordinary helical edge modes (blue curves), and second-order zero-dimensional corner states (red curves). **b,** Spectrum of the density of states at $I_1$. The resolved spectrum includes neither a corner nor an edge mode, and the crystal is topologically trivial. **c,** Spectrum of the density of states at the intensity $I_2$ above threshold, evidencing the presence of both edge and corner states. **d,** When the intensity is further increased to $I_3$, the band gap becomes larger.

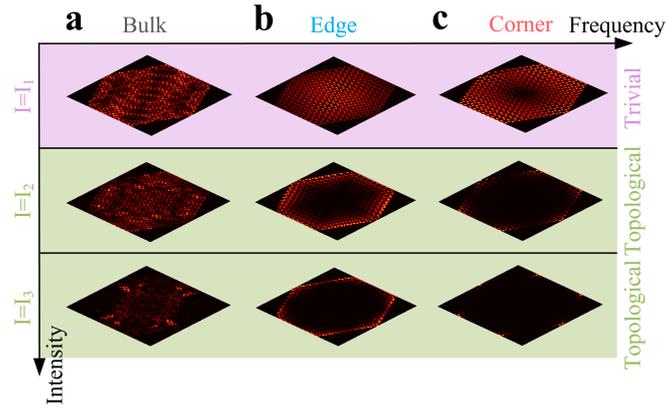

**Fig. 4: Intensity controlled edge and corner states in a non-linear SOTI. a,** Mode profile of an ordinary bulk Bloch state as the intensity level is increased. **b,** Same as column a, but for one of the corresponding edge modes. At the intensity levels ($I_1$), corresponding to a trivial topology, the mode actually lives in the bulk. When the intensity level is enhanced to $I_2$, corresponding to a non-trivial topology, the mode localizes at the perimeter of the sample, creating an edge mode. Increasing the intensity level further to $I_3$ increases the localization of the edge mode. **c,** Same as panel b but for a zero-dimensional corner state.

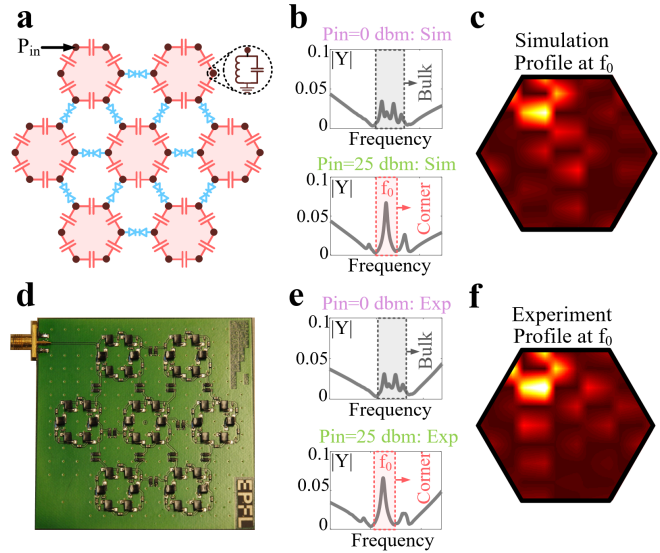

**Fig. 5, Experimental demonstration of self-induced nonlinear SOTIs**, **a,** We map the proposed tight binding lattice into a circuit, consisting of *LC* resonators capacitively coupled to each other via linear capacitances (red) and back-to-back nonlinear varactor diodes (blue). **b,** (Top) Numerically calculated input admittance spectrum of the circuit when it is excited at very low excitation power, corresponding to a trivial topology. The spectrum includes multiple resonances, which correspond to bulk modes. When the excitation intensity is enhanced, a dominant resonance appears in the spectrum (bottom), corresponding to the topological 0D corner state. **c,** Numerically calculated mode profile of the topological corner state. **d,** Picture of the prototype. **e,f,** Measurements corresponding to panels b-c.

**Acknowledgments**

This work was supported by the Swiss National Science Foundation (SNSF) under Grant No. 172487.